\documentstyle[preprint,aps,prl,epsf]{revtex}

\begin{document}

\draft

\title{ Progress in Monte Carlo Calculations of Fermi Systems:
Normal Liquid $^{\bf 3}$He }

\author{J. Casulleras and J. Boronat}

\address{Departament de F\'{\i}sica i Enginyeria Nuclear,
Campus Nord B4-B5, \protect\\ Universitat Polit\`ecnica de Catalunya,
E-08034 Barcelona, Spain}
\date{\today}

\maketitle

\begin{abstract}
The application of the diffusion Monte Carlo method to a strongly
interacting Fermi system as normal liquid $^3$He is explored.
We show that the fixed-node method together with the released-node
technique and a systematic method to analytically improve the nodal
surface constitute an efficient strategy to improve the calculation up to
a desired accuracy. This methodology shows unambiguously that
backflow correlations, when properly optimized, are enough to
generate an equation of state of liquid $^3$He in excellent agreement with
experimental data from equilibrium up to freezing.
\end{abstract}



Liquid $^{3}$He has been for many years a benchmark in the field of
quantum-many body physics. The Fermi statistics of its atoms, combined
with the strong correlations induced by the hard core of their interatomic
potential, has turned it into a paradigm of strongly-correlated Fermi
systems. At zero temperature, an approximate microscopic description has been
achieved by means of variational methods, both for the Fermi liquid
$^{3}$He \cite{kalos,fanto,lhui} as well as for its bosonic counterpart
liquid $^{4}$He \cite{schmidt}. From a Monte Carlo viewpoint, the quantum
many-body problem can be tackled in a more ambitious way with the aid of
the Green's function Monte Carlo (GFMC) and the diffusion Monte Carlo
(DMC) methods \cite{abini}. The variational wave function can
be used as an input for the Monte Carlo method which, for boson systems,
is able to solve the Schr\"{o}dinger equation of the $N$-body system
providing exact results. In Fermi systems as liquid $^{3}$He, the
exactness of the method is lost due the involved sign problem that makes a
straightforward interpretation of the wave function not possible.

The cancellation methods developed up to now to solve this intricate
problem have proved their efficiency in model problems or with very few
particles but become unreliable for real many-body systems \cite{corr}. In
the meantime, the approximate fixed-node (FN) \cite{fnod} method has
become a standard tool. In the FN-DMC method, the antisymmetry is
introduced in the trial wave function used for importance sampling
imposing its nodal surface as a boundary condition. This approach provides
upper bounds to the exact eigenvalues, the quality of which is related to
the accuracy of the nodal surface of the trial wave function. The main
drawback of the FN-DMC method is the lack of control over the influence of
the imposed nodal surface on the results obtained, not to say the
impossibility of properly correcting for such effect. In the present
Letter, we come back to this problem using the FN-DMC as a main approach
but crucially combined with two auxiliary methods: the released-node (RN)
estimation technique \cite{cepe} and an analytical method able to enhance
the quality of any given nodal surface. The combination of the above
methods provides information on the bias due to the imposed nodal surface
and a procedure which can evaluate and
bring down, in principle to any arbitrary required precision, this influence. 
This
complete program has been applied to the study of liquid $^{3}$He
bringing, as we will show, the systematic error under control and down to
levels below the current statistical errors. This has
allowed a very accurate microscopic calculation of the equation of state
of liquid $^3$He, including a prediction for the negative-pressure region
and the spinodal density.

DMC \cite{moro} and GFMC \cite{pano} calculations have
provided up to now the best upper bounds to the ground-state energy of
liquid $^3$He .  These calculations have unambiguously shown the relevance
of the Feynman-Cohen-type backflow correlations in the improvement of the
energy. However, the energy gain appears too small to recover the
experimental data and, what is more conclusive, the density dependence of
the pressure which stresses the curvature of the equation of state shows
clear differences with the experiment. A conclusion that naturally emerged
from those results was that the nodal surface, originated by backflow
correlations, is not accurate enough and probably new state-dependent
correlations ought to be considered. Our present results prove that
backflow correlations, when properly optimized, effectively do provide very
accurate nodal surfaces.

In the DMC method, the Schr\"{o}dinger equation written in imaginary time
is translated into a diffusion-like differential equation which can be
stochastically solved in an iterative procedure. Specific information on
the implementation of the DMC method is given in Ref. \cite{boro}. As far
as the FN framework \cite{fnod} is concerned, the choice of the trial wave
function $\psi ({\textbf {R}})$ used for importance sampling is a key point.
The simplest model is the Jastrow-Slater wave function 
\begin{equation}
\psi= \psi_{\rm J} \, D_{\uparrow} D_{\downarrow}   \ ,
\label{jassimple}
\end{equation}
with $\psi_{\rm J} = \prod_{i<j} f(r_{ij}) $ a Jastrow wave function and
$D_{\uparrow}$ ($D_{\downarrow}$) a Slater determinant of the spin-up
(spin-down) atoms with single-particle orbitals $\varphi_{\alpha_i}
({\bf r}_j)= \exp (i \, {\bf k}_{\alpha_i}\cdot {\bf r}_j)$. 

In this variational description (\ref{jassimple}), the dynamical
correlations induced by the interatomic potential are well modelled by the
Jastrow factor, and the statistical correlations, implied by the
antisymmetry, are introduced with a Slater determinant of plane waves
which is the exact wave function of the free Fermi sea. The two factors
account well for the dynamical correlations and the Fermi statistics when
these effects are independently considered but their product is only a
relatively poor approximation for a strongly correlated Fermi liquid. It
is well known, from previous variational and GFMC/DMC calculations
\cite{moro,pano}, that a significant improvement on the Jastrow-Slater
model is achieved by introducing backflow correlations in $\varphi
_{\alpha _{i}}({\textbf {r}}_{j})$, a name which is taken from the
Feynman-Cohen famous work on the microscopic description of the
phonon-roton spectrum in liquid $^{4}$He \cite{feyn}.

At this point, it becomes essential to set up a method for analytically
enhancing a given model. Such a procedure is already contained in the
imaginary-time Schr\"{o}dinger equation.  Let us consider a time-dependent
wave function $\phi({\bf R},t)$, with $\phi({\bf R},t=0)=\psi({\bf R})$
the initial guess for the trial wave function, satisfying 
\begin{equation}
-\frac{\partial \phi({\bf R},t)}{\partial t} = H \phi({\bf R},t)  \ .
\label{timed} 
\end{equation}          
A natural choice for a more accurate trial wave function is obtained
solving Eq. (\ref{timed}) at first order in $t$. Near the nodes, which is
the relevant region to our purposes, one readily
captures the main correction to the original $\psi_{\rm A}({\bf R}) \equiv
D_{\uparrow} D_{\downarrow}$ in the form $\phi({\bf R},t)=\psi_{\rm
J}({\bf R}) \psi_{\rm A}(\tilde{{\bf R}})$ with $\tilde{\bf R}= {\bf
R}(t)$ and $\partial {\bf R}/\, \partial t = D {\bf F}_{\rm J}({\bf R})$,
$ {\bf F}_ {\rm J}({\bf R})$ being the drift force coming from the Jastrow
wave function $\psi_{\rm J}({\bf R})$ and $D=\hbar^2/(2m)$. In this form,
the new nodal surface is described by the original antisymmetric wave
function but with arguments that are shifted due to the effect of
dynamical correlations. It is worth noting that this approach generates
the Feynman-Cohen backflow in $\varphi_{\alpha_i}({\bf r}_j)$ ($\psi_{\rm
A}^{\rm BF}({\bf R})$) as a first order correction to the plane-wave
orbitals. Recursively, entering with $\psi_{\rm A}^{\rm BF}({\bf R})$ the
next order correction can be analytically
obtained (see Eq. \ref{backt}).                                                        

Finally,  once a specific model for the nodal surface has been chosen it
is  necessary to establish a method to test its quality. This can be
accomplished by means of the released-node technique \cite{cepe}. In the
RN approach a superposition of a small boson component in the wave
function is allowed, with the primary effect of resetting the nodal
surface to the exact position. This is technically accomplished by
introducing a positively-defined guiding wave function $\psi_{\rm
g}({\textbf {R}})$  so that the walkers are not confined into a region of
definite sign of $\psi ({\textbf {R}})$. The basic requirements on
choosing $\psi _{\rm g}({\textbf {R}})$ are twofold: proximity to $|\psi
({\textbf {R}})|$ away from the nodal surface and being positive-defined
at  the nodes. The choice we have made is
\begin{equation} 
\psi_{\rm g}({\bf R}) = ( \psi({\bf R})^2 + a^2)^{1/2}   
\ , 
\label{psig} 
\end{equation} 
$a$ being a parameter which controls the
crossing frequency. Other different choices can be considered but the
specific details of $\psi _{\rm g}({\textbf {R}})$ are not relevant
since the RN energy is calculated projecting out its antisymmetric
component. The use of $\psi _{\rm g}({\textbf {R}})$ does not
introduce any systematic bias in the RN energies, which approach the exact
eigenvalue when $t_{\rm r}\rightarrow \infty$ , $t_{\rm r}$ being the maximum
allowed lifetime after the first crossing. However, the variance of the
energy grows exponentially with $t_{\rm r}$ due to the boson component, and
thus in general the asymptotic value cannot be obtained. In contrast, what is
straightforwardly available is the slope of the energy versus $t_{\rm r} $ at
$t_{\rm r}  \rightarrow 0$, which provides a direct measure of the quality of the
input nodal surface (the true antisymmetric ground-state wave function
would generate a zero slope), and constitutes a means of comparing
different trial wave functions. In particular, it provides feedback
information on whether the next analytical correction to $\psi_{\rm A}^{\rm
BF}({\bf R})$ is necessary.

We have applied all the above methodology to the study of normal liquid
$^3$He at zero temperature. 
The results reported have been obtained with $N=66$ particles,
but we have made size checks using also $N=54$ and $N=114$. In Fermi
systems, the kinetic energy includes statistical contributions that show
an  oscillating behavior with $N$. We have observed that this behavior  
follows very closely that of a discretized Fermi-gas energy, a fact that
could be expected since such a term appears explicitly in the  local
kinetic energy. It is worth noticing that the case $N=66$ is specially
well suited for MC calculations as the correction amounts only 0.015 K.

As in previous calculations, we use a
short-ranged backflow in the form $\varphi_{\alpha_i}({\bf r}_j)= \exp (i
{\bf k}_{\alpha_i} \cdot \tilde{{\bf r}}_j^{\rm BF} )$, with
\begin{equation} 
\tilde{{\bf r}}_j^{\rm BF} = {\bf r}_j + \lambda_{\rm B}
\sum_{k \neq j} \eta(r_{jk}) \, {\bf r}_{jk}  \ , 
\label{back1}
\end{equation} 
and $\eta(r)= \exp ( - ( (r-r_{\rm B})/\, \omega_{\rm B})^2
)$. The two-body correlation factor has been chosen of McMillan type,
$f(r)=\exp(-0.5\, (b/\, r)^5)$, and the pairwise HFD-B(HE) Aziz potential
\cite{aziz}, which has proved high accuracy in liquid $^4$He calculations
\cite{boro},  has modelled the atomic interactions. At the experimental
equilibrium density $\rho_0^{\rm expt}=0.273\ \sigma^{-3}$ ($\sigma=2.556$
\AA), we have started the calculation with $b=1.15\ \sigma$ and the
backflow parameters  optimized in Ref. \cite{pano} ($\lambda_{\rm
B}=0.14$, $r_{\rm B}=0.74\ \sigma$, $\omega_{\rm B}=0.54\ \sigma$).
With this initial set of parameters, the results obtained are clearly
biased by the trial wave function.
Even though the RN approach \cite{cepe} corrects
numerically the shortcomings of $\psi({\bf R})$ and, in some applications,
allows for an exact estimation of the eigenvalue, this is not the
case for liquid $^3$He. In Fig. 1, the RN energies  as a function of the
released times are shown for the cases $\lambda_{\rm B}=0$ and  
$\lambda_{\rm B}=0.14$.
As one can see, at small imaginary times the RN method
reveals the presence of corrections to the FN energies but a common
asymptotic regime is far beyond the scope of the available MC data. 

The next step then was looking for the next order correction to backflow
correlations, as well as for a possibly better set of backflow parameters.
We have found that the ones we were using correspond
to a local minimum of the FN energy, and
that a narrower but deeper minimum exists with $\lambda_{\rm B}=0.35$ and
$r_{\rm B}$ and  $\omega_{\rm B}$ unchanged. The resulting energy versus 
released time is also plotted in Fig. 1. 
The relation of initial slopes, $1:0.27:0.016$ for $\lambda_{\rm
B}=0,\ 0.14,\ 0.35$, provides information on the accuracy of $\psi({\bf
R})$. In the optimal case, $\lambda_{\rm B}=0.35$, the
slope is practically inexistent and the energy correction would be $\lesssim
0.01$ K if the asymptotic regime could be reached. In order to get
additional evidence on the size of this correction, and as a closing
checkmark of the reliability of our results, we have included corrections
to the backflow trial wave function using the analytical method 
previously described. It can be shown that these new terms incorporate 
explicit
three-body correlations in $\varphi_{\alpha_i} (\tilde{{\bf r}}_j)$ of the
form
\begin{equation} 
\tilde{{\bf r}}_j^{\rm BFT} =
\tilde{{\bf r}}_j^{\rm BF} + \lambda_{\rm BT} \sum_{k \neq j} \eta(r_{jk})
\, ({\cal F}_j - {\cal F}_k )  \ , 
\label{backt} 
\end{equation} 
with
${\cal F}_i = \sum_{l \neq i} \eta(r_{il}) \, {\bf r}_{il}$. We have
carried out a FN-DMC calculation with this new trial wave function at
$\rho_0^{\rm expt}$ and the result for the energy correction has been
found  $< 0.01$ K. Both this analytical check and the numerical
findings provided by the RN method point out the excellent description
that backflow correlations make of the nodal surface in liquid $^3$He.

The FN energies with
$\lambda_{\rm B}=0$ (no backflow), $\lambda_{\rm B}=0.14$, and
$\lambda_{\rm B}=0.35$ are reported in Table I, together with the
corresponding kinetic energy obtained as the difference between the total
energy and a pure estimation \cite{pure} of the potential energy. 
The comparison with
the experimental energy \cite{yang}, also contained in the table, shows
the successive improvement of the FN-DMC result until an excellent 
agreement with $\lambda_{\rm B}=0.35$. Concerning the kinetic energy, a
sizeable difference between theory and experiment \cite{scat} survives, a
fact that has been generally attributed to long-range wings in the
high-$q$ inelastic response that are difficult to incorporate effectively
in the experimental analysis.

The FN-DMC calculation has been extended to a wide range of densities
ranging from the spinodal point up to a maximum value $\rho=0.403\
\sigma^{-3}$, located near to the experimental freezing density $\rho_{\rm
f}^{\rm expt}=0.394\ \sigma^{-3}$. We have used $N=114$ only at the
highest density and below that $N=66$ have proved to be accurate enough.
Among the three variational parameters entering in the backflow wave
function (\ref{back1}) only $\lambda_{\rm B}$ shows a density dependence
which is nearly linear in the range studied ($\lambda_{\rm B}=0.42$ at
$\rho=0.403\ \sigma^{-3}$). The results are displayed in Fig. 2 in
comparison with the experimental data of Ref. \cite{yang}. The solid line
in the same figure is a third-degree polynomial fit to our data with
$\chi^2/\nu = 1.2$. According to this fit, the equilibrium density is
$\rho_0=0.274(1)\ \sigma^{-3}$ and the energy at this density $(E/N)_0 =
-2.464(7)$ K, in close agreement with experimental data. 

The quality of the equation of state is even more stressed by looking at
its derivatives. In Fig. 3, the behavior of the pressure and the sound
velocity with the density is shown in comparison with experimental data
from Refs. \cite{yang,so1,so2}. The  theoretical prediction for both
quantities, derived form the polynomial fit to $E/N(\rho)$ (Fig.
1), shows again an excellent agreement with the experimental data from
equilibrium up to freezing. The sound velocity that at $\rho_0^{\rm expt}$
is $c=182.2(6)$ m$/\,$sec, in close agreement with the experimental value
$c^{\rm expt}=182.9$ m$/\,$sec \cite{so1}, goes  down to zero at the
spinodal point. The location of this point has been previously obtained
both from extrapolation of experimental data at positive pressures
\cite{bali} and from density-functional theories \cite{jeze}. The present
microscopic  calculation allows for an accurate calculation, free from
extrapolation uncertainties, that locates the spinodal point at a density
$\rho_{\rm s}=0.202(2)\ \sigma^{-3}$ corresponding to a negative pressure
$P_{\rm s}=-3.09(20)$ atm, much closer to the equilibrium than in liquid $^4$He
where $P_{\rm s}=-9.30(15)$ atm \cite{jesus}.

In conclusion, we have analyzed the possibilities of the diffusion Monte
Carlo method in the study of Fermi systems.  The FN-DMC method,  combined
with the RN technique to evaluate the effect of the nodal surface model
used in the trial function, and with a systematic method to analytically
improve the nodal surface, constitutes a closed loop able  to improve the
quality of the antisymmetric wave function up to a required precision. The
fact that this general approach has allowed to deal with a strongly
interacting Fermi liquid suggests that it could be
also useful for tackling other Fermi systems. We have applied  this
methodology to the study of liquid $^3$He.  The effect of corrections
beyond the backflow terms has been evaluated and found to be less than
0.01 K. The equation of state so obtained presents an accuracy comparable
to the result obtained in  bosonic liquid $^4$He. The precision proved by
the method allows for posterior studies as the characterization of
spin-polarized liquid $^3$He. Preliminary calculations for the
fully-polarized phase at  $\rho_0^{\rm expt}$ indicate a less binded
system with an energy $E/N=-2.22(4)$ K.       

This research has been partially supported by DGES (Spain) Grant N$^0$
PB96-0170-C03-02. We also acknowledge the supercomputer facilities
provided by the
CEPBA.

\begin{table}

\caption{FN-DMC total and kinetic energies for liquid $^3$He at
$\rho_0^{\rm expt}$ as a function of the backflow parameter $\lambda_{\rm
B}$. The experimental values for the total and kinetic energies are taken form
Ref. \protect\cite{yang} and Ref. \protect\cite{scat}, respectively.}

\begin{tabular}{lcc}
   &    $E/N$ (K)   & $T/N$ (K) 
\\ \tableline
$\lambda_{\rm B}=0$       &  $-2.128 \pm 0.015$     & $12.603 \pm 0.031$  \\ 
$\lambda_{\rm B}=0.14$    &  $-2.330 \pm 0.014$     & $12.395 \pm 0.035$  \\
$\lambda_{\rm B}=0.35$    &  $-2.477 \pm 0.014$   & $12.239 \pm 0.030$  \\
 Expt                     &  -2.473             &   $8.1 \pm 1.7$      \\
\end{tabular}

\end{table}

\begin{figure}
\caption{Relesead-node energies as a function of the released time. Open
circles, triangles and full circles correspond to $\lambda_{\rm B}=0$ (no
backflow), $\lambda_{\rm B}=0.14$, and $\lambda_{\rm B}=0.35$,
respectively. The lines are linear fits to the MC data.}
\begin{center}
\epsfxsize=18cm  \epsfbox{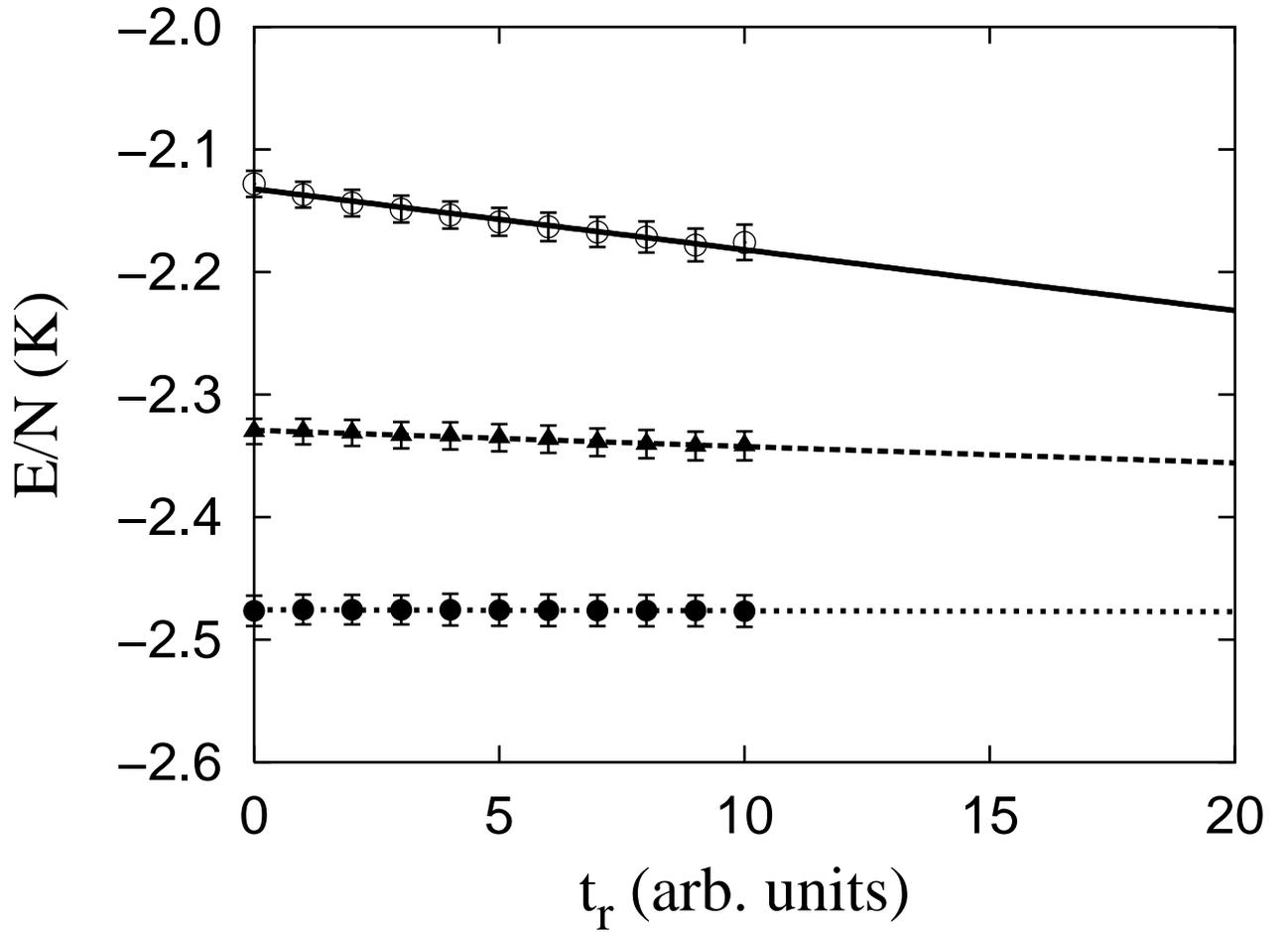}
\end{center}                                      
\end{figure}                                                           

\begin{figure}
\caption{Energy per particle of normal liquid $^3$He as a function of the
density. The full circles are the FN-DMC results (the error bars are depicted
only when larger than the size of the symbol), and the open circles
are experimental data from Ref. \protect\cite{yang}. The line is a 
polynomial fit to the MC data.}

\begin{center}
\epsfxsize=17cm  \epsfbox{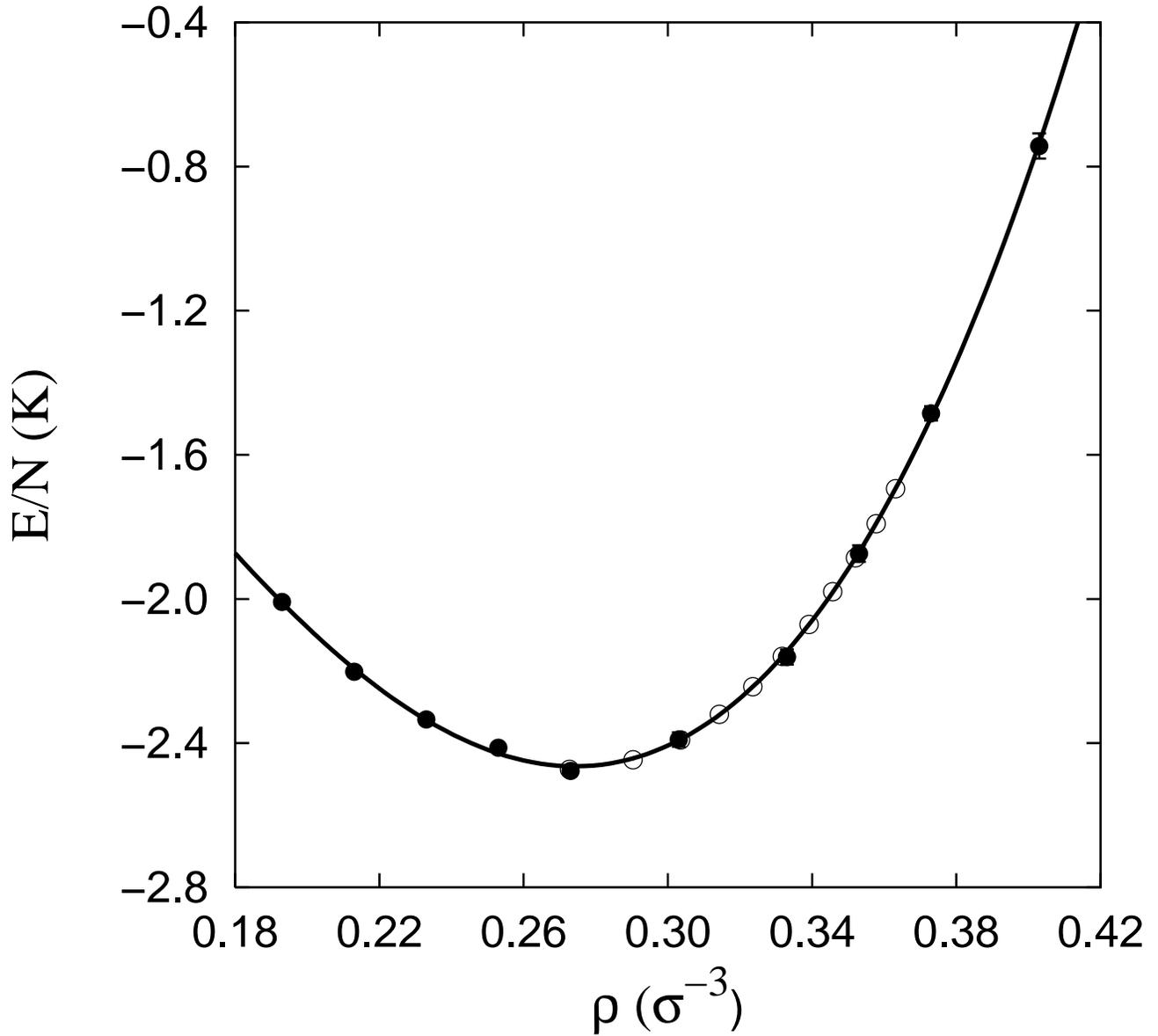}
\end{center}                                      
\end{figure}

\begin{figure}
\caption{Pressure and sound velocity as a function of the density. The
lines are the FN-DMC results and the circles, triangles and squares are 
experimental data from Refs. \protect\cite{yang}, \protect\cite{so1}, and
\protect\cite{so2}, respectively.}

\begin{center}
\epsfxsize=17cm  \epsfbox{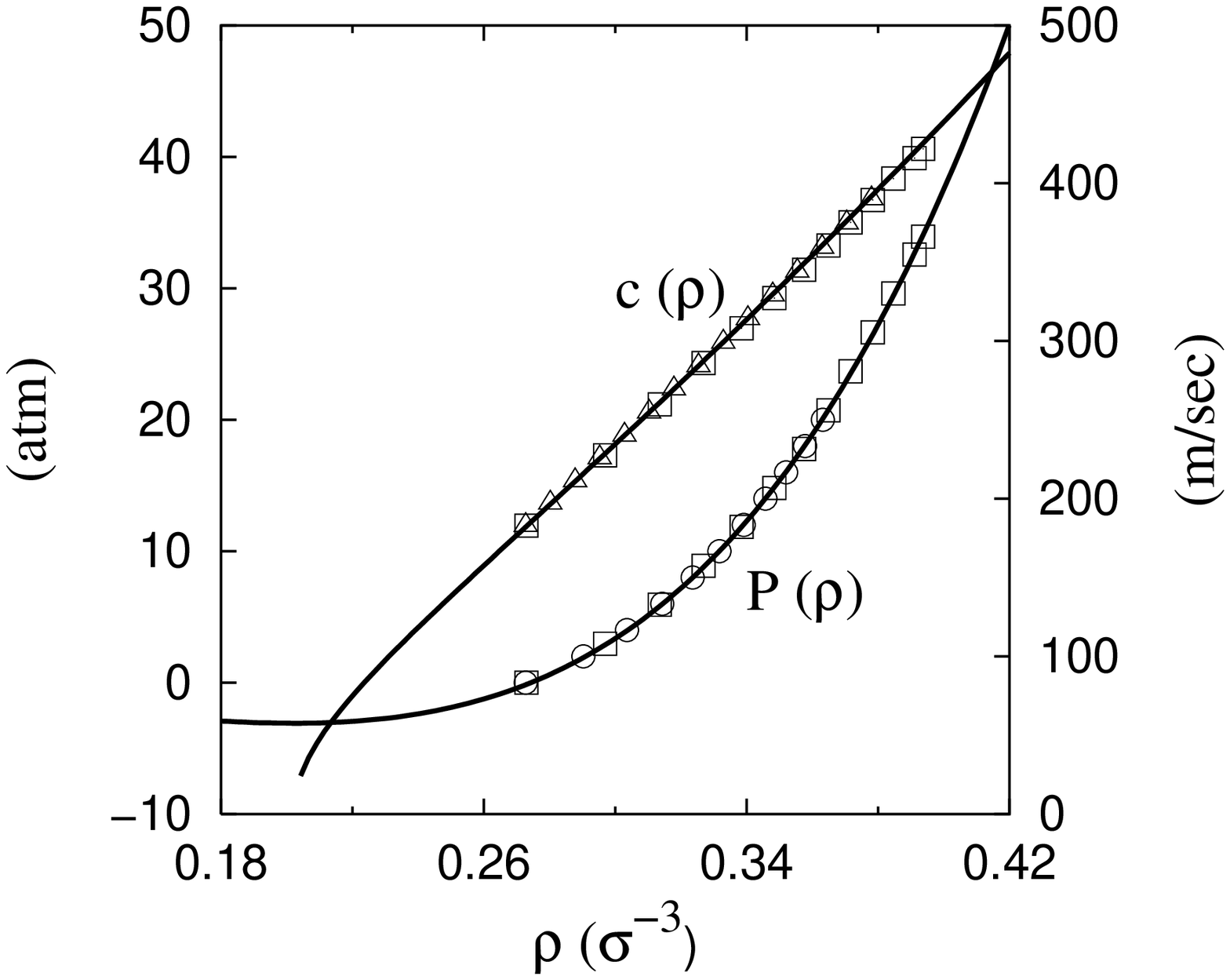}
\end{center}                        
\end{figure}


\begin{references}

 
\bibitem{kalos} K. E. Schmidt, M. A. Lee, M. H. Kalos, and G. V. Chester, 
Phys. Rev. Lett. {\bf 47}, 807 (1981).                         
                          
\bibitem{fanto} M. Viviani, E. Buend\'\i a, S. Fantoni, and S. Rosati, Phys. 
Rev. B {\bf 38}, 4523 (1988).      

\bibitem{lhui} J. P. Bouchaud and C. Lhuillier, Z. Phys. B {\bf 75}, 283 
(1989).

\bibitem{schmidt} K. E. Schmidt, M. H. Kalos, M. A. Lee, and G. V. Chester, 
Phys. Rev. Lett. {\bf 45}, 573 (1980).  
                     
\bibitem{abini} B. L. Hammond, W. A. Lester Jr., and P. J. Reynolds,  
{\it Monte Carlo Methods in Ab Initio Quantum Chemistry} (World
Scientific, Singapore, 1994).                                                                 

\bibitem{corr} S. Zang and M. H. Kalos, Phys. Rev. Lett. {\bf 67}, 
3074 (1991).      

\bibitem{fnod} P. J. Reynolds, D. M. Ceperley, B. J. Alder, and 
W. A. Lester, J. Chem. Phys. {\bf 77}, 5593 (1982). 
                                 
\bibitem{cepe} D. M. Ceperley and  B. J. Alder, Phys. Rev. Lett. {\bf 45}, 
566 (1980).
   
\bibitem{moro} S. Moroni, S. Fantoni, and G. Senatore, Phys. Rev. B {\bf
52}, 13 547 (1995).  

\bibitem{pano} R. M. Panoff and J. Carlson, Phys. Rev. Lett. {\bf 62}, 
1130 (1989). 

\bibitem{boro} J. Boronat and J. Casulleras, Phys. Rev. B {\bf 49}, 
8920 (1994).   

\bibitem{feyn} R. P. Feynman and M. Cohen, Phys. Rev. {\bf 102}, 1189
(1956). 

\bibitem{aziz} R. A. Aziz, F. R. W. McCourt, and C. C. K. Wong,  
Mol. Phys. {\bf 61}, 1487 (1987).
                      
\bibitem{pure} J. Casulleras and J. Boronat, Phys. Rev. B {\bf 52}, 3654
(1995).
                      
\bibitem{yang} R. De Bruyn Ouboter and C. N. Yang, Physica B {\bf 144}, 
127 (1987).              

\bibitem{scat} P. E. Sokol, K. Sk\"old, D. L. Price, and R. Kleb, Phys.
Rev. Lett. {\bf 54}, 909 (1985).

\bibitem{so1} R. A. Aziz and R. K. Pathria, Phys. Rev. A {\bf 7}, 809
(1973).

\bibitem{so2} J. C. Wheatley. Rev. Mod. Phys. {\bf 47}, 415 (1975).

\bibitem{bali} F. Caupin, P. Roche, S. Marchand, and S. Balibar, J. Low
Temp. Phys. {\bf 113}, 473 (1998).

\bibitem{jeze} D. M. Jezek, M. Pi, and M. Barranco, Phys. Rev. B, in press.

\bibitem{jesus} J. Boronat, J. Casulleras, and J. Navarro, Phys. Rev. B
{\bf 50}, 3427 (1994).




\end{references}
\end{document}